\documentclass[journal]{IEEEtran}
\usepackage{cite}
\usepackage{amsmath,amssymb,amsfonts}
\usepackage{algorithmic}
\usepackage{graphicx}
\usepackage{textcomp}

\usepackage{multicol}
\usepackage{graphicx}
\usepackage{amsmath}
\usepackage{mathrsfs}
\usepackage[table,xcdraw]{xcolor}
\usepackage{multirow}
\usepackage{array}
\usepackage{booktabs}
\usepackage{subcaption}
\usepackage{adjustbox}
\usepackage{threeparttable}

\begin{document}
%
\title{Deep Learning-based End-to-end Diagnosis System for Avascular Necrosis of Femoral Head}

%
%
%

\author{
	Yang~Li,
	Yan~Li,
	Hua~Tian
	\thanks{This manuscript has been accepted by IEEE Journal of Biomedical and Health Informatics. After publication, it will be found at https://doi.org/10.1109/JBHI.2020.3037079. Yang Li and Yan Li contributed equally to this paper. \textit{(Corresponding author: Hua Tian)}}
	\thanks{Yang Li (email: liyangdr@bjmu.edu.cn) and Hua Tian (email: tianhua@bjmu.edu.cn) are with the Orthopaedics department of Peking University Third Hospital and Engineering Research Center of Bone and Joint Precision Medicine, Ministry of Education, Beijing, 100191, China}
	\thanks{Yan Li is with the Department of Electrical \& Computer Engineering, University of Toronto, ON M5S 3G4, Canada (email: yyaann.li@mail.utoronto.ca)}
}

%
%

\markboth{IEEE JOURNAL OF BIOMEDICAL AND HEALTH INFORMATICS}%
{LI \MakeLowercase{\textit{et al.}}: Deep Learning-based End-to-end Diagnosis System for Avascular Necrosis of Femoral Head}
%



\maketitle

\begin{abstract}
As the first diagnostic imaging modality of avascular necrosis of the femoral head (AVNFH), accurately staging AVNFH from a plain radiograph is critical yet challenging for orthopedists. Thus, we propose a deep learning-based AVNFH diagnosis system (AVN-net). The proposed AVN-net reads plain radiographs of the pelvis, conducts diagnosis, and visualizes results automatically. Deep convolutional neural networks are trained to provide an end-to-end diagnosis solution, covering tasks of femoral head detection, exam-view identification, side classification, AVNFH diagnosis, and key clinical notes generation. AVN-net is able to obtain state-of-the-art testing AUC of $0.97$ ($95\%$ CI: $0.97-0.98$) in AVNFH detection and significantly greater F1 scores than less-to-moderately experienced orthopedists in all diagnostic tests ($p$\textless$0.01$). Furthermore, two real-world pilot studies were conducted for diagnosis support and education assistance, respectively, to assess the utility of AVN-net. The experimental results are promising. With the AVN-net diagnosis as a reference, the diagnostic accuracy and consistency of all orthopedists considerably improved while requiring only 1/4 of the time. Students self-studying the AVNFH diagnosis using AVN-net can learn better and faster than the control group. To the best of our knowledge, this study is the first research on the prospective use of a deep learning-based diagnosis system for AVNFH by conducting two pilot studies representing real-world application scenarios. We have demonstrated that the proposed AVN-net achieves expert-level AVNFH diagnosis performance, provides efficient support in clinical decision-making, and effectively passes clinical experience to students.
\end{abstract}


\begin{figure}[t]
	\centering
	\begin{subfigure}[b]{\columnwidth}
		\centering
		\includegraphics[width=0.22\columnwidth]{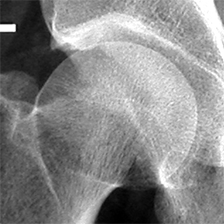}
		\hspace{0.05\columnwidth}
		\includegraphics[width=0.22\columnwidth]{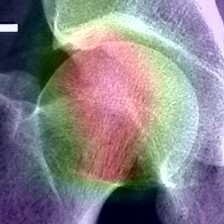} 
		\caption{AVNFH absence: No findings}
	\end{subfigure}
	
	\vspace{1ex}
	\begin{subfigure}[b]{\columnwidth}
		\centering
		\includegraphics[width=0.22\columnwidth]{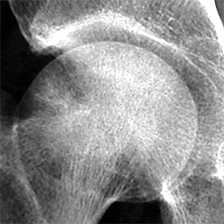}
		\hspace{0.05\columnwidth}
		\includegraphics[width=0.22\columnwidth]{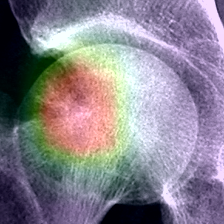} 
		\caption{AVNFH stage II: Sclerotic and cystic changes}
	\end{subfigure}
	
	\vspace{1ex}
	\begin{subfigure}[b]{\columnwidth}
		\centering
		\includegraphics[width=0.22\columnwidth]{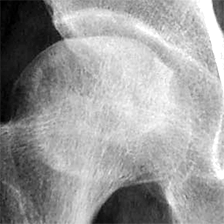} 
		\hspace{0.05\columnwidth}
		\includegraphics[width=0.22\columnwidth]{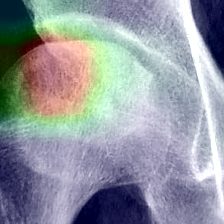}
		\caption{AVNFH stage III: Subchondral flattening/collapse}
	\end{subfigure}
	
	\vspace{1ex}
	\begin{subfigure}[b]{\columnwidth}
		\centering
		\includegraphics[width=0.22\columnwidth]{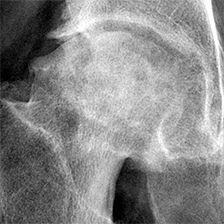} 
		\hspace{0.05\columnwidth}
		\includegraphics[width=0.22\columnwidth]{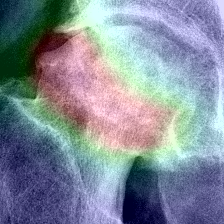}
		\caption{AVNFH stage IV: FH and acetabular deformation}
	\end{subfigure}
	
	\caption{Examples of different AVNFH stages (left), visualized diagnosis with class activation map (right), and generated clinical notes (captions).}
	\label{fig:1}
\end{figure}

%
\IEEEpeerreviewmaketitle

\section{Introduction}
\IEEEPARstart{A}{vascular} necrosis of the femoral head (AVNFH) is characterized as progressive pain and disabling degeneration of the hip joint that typically affects the active, middle-aged population~\cite{mont1995non}~\cite{petek2019osteonecrosis}. In the United States, over 20,000 new AVNFH patients are diagnosed annually~\cite{moya2015current}, and a high probability of bilateral involvement (73\%) has been reported~\cite{boettcher1970non}. In China, the number of AVNFH patients has reached approximately 8 million as of 2015~\cite{zhao2015prevalence}, and the number of new cases for each year is around 100,000~\cite{zhao2012clinical}. Note that early presentation of AVNFH is typically asymptomatic and painless~\cite{sen2009management}. But, over 70\% of asymptomatic cases can transform to symptomatic and femoral head (FH) collapse which ultimately requires hip arthroplasty~\cite{hernigou2004fate, bradway1993natural, jergesen1997natural}, and the mean interval between the first symptoms and the FH collapse is only 12 months~\cite{hernigou2004fate}. Although there is no consensus on the etiology of AVNFH, a range of risk factors have been identified; thus, various interventions have been developed to alleviate the pain and delay progression~\cite{petek2019osteonecrosis}\cite{moya2015current}\cite{sen2009management}. However, the success of treatments is highly dependent on the stage at which care is initiated \cite{mont1995non}. Therefore, timely and accurate staging is essential to optimizing AVNFH treatments. 

Clinically, the first diagnostic imaging modality for AVNFH is radiography due to its cost, accessibility, and efficiency advantages. If more imaging evidences are needed, extra exams will be performed, such as computerized tomography (CT) or magnetic resonance imaging (MRI). Radiographs from both anteroposterior (AP) and frog-leg lateral (FL) views are typically required to obtain a comprehensive diagnosis because AP and FL views contain complementary areas of the hip joint~\cite{mont1995non}. The Ficat staging system, a widely used AVNFH diagnosis standard mainly based on radiographs, provides practical guidance for classifying AVNFH into five stages with severity increasing from stage 0 (normal imaging) to stage IV~\cite{ficat1985idiopathic}. However, the modest changes in the early stage of AVNFH (e.g., Fig.~\ref{fig:1}b) pose significant challenges to diagnostic sensitivity~\cite{stoica2009imaging}, which results in low inter-/intra-reader agreements~\cite{kay1994interobserver}. Therefore, clinicians usually require years of training and practice to perform accurate diagnoses. 

With rapid advancements in deep learning (DL) methodologies, deep convolutional neural networks (DCNN) have demonstrated exceptional capabilities in solving diagnostic problems. Using annotated data, some DCNN algorithms have reached the performance of human experts in various fields, e.g., in ophthalmology~\cite{gulshan2016development}\cite{li2018efficacy}, dermatology~\cite{esteva2017dermatologist}\cite{haenssle2018man}, and orthopedics~\cite{urakawa2019detecting}\cite{chung2018automated}. Unfortunately, only one study to date has attempted to approach AVNFH diagnosis by leveraging DL methodologies which possibly because of two reasons. First, there is no publicly available dataset with high quality of annotations for AVNFH diagnosis. More importantly, the nature of AVNFH poses a number of challenges to DL algorithms and hence makes this task more difficult. Compared to some thorax disease (e.g., pneumonia), the lesion area of AVNFH is highly localized relative to the size of radiographs; thus, the majority proportion of the information on a pelvic radiograph would be irrelevant to diagnostic decision-making. To improve the performance, it is common to manually crop the region-of-interest (ROI) from a radiograph before training. However, this process inevitably complicates the data preparation and impedes the application of the DL model. In addition, in most studies of DL in medicine, discussions typically stop at performance comparisons between algorithms and experts. No further analysis has been provided to demonstrate how well DL models perform in real-world scenarios.

To address these issues, we propose the AVN-net, a fully automated AVNFH diagnosis system based on DCNNs and plain pelvic radiographs. AVN-net comprises five functional modules to perform the FH detection, exam view identification, side classification, AVNFH diagnosis, and key clinical note generation tasks in an end-to-end fashion. The FH detection module extracts the ROI (i.e., the FH) from the input radiograph. Then, four downstream modules are trained using the detected FH data to perform the remaining tasks. Experimental results demonstrate that the proposed AVN-net can achieve state-of-the-art testing performance and significantly outperforms less-to-moderately experienced orthopedists.

We further performed two pilot studies to investigate the utility of AVN-net in real-world application scenarios. In the first study, AVN-net was used to provide orthopedists with a ``second opinion" in AVNFH staging. Here, we compared the diagnostic performance and time consumption of orthopedists with and without AVN-net to assess its effectiveness in supporting clinical decision making. In the second study, AVN-net was applied as an experienced teaching assistant to help medical students self-learn AVNFH diagnosis because we believe the quality of self-study plays a critical role in medical education due to increasingly limited educational resources. We tested the performance of two groups of students (i.e., self-study with and without AVN-net) in a closed-book test after the same learning time to evaluate the usefulness of AVN-net. Both studies yielded promising results and confirmed that AVN-net could achieve expert-level performance in AVNFH diagnosis and provide effective support in clinical decision making and medical education.

The remainder of this paper is organized as follows. Section 2 reviews progress made in disease diagnosis and object detection with DL. Our primary contribution is also summarized in Section 2. Section 3 introduces all methods adopted in this study. Results are discussed in Section 4. Section 5 describes the designs and results of pilot studies, followed by discussions in Section 6. Finally, conclusions are presented in Section 7.

\section{Related work}

\subsection{Diagnosis with deep learning}
Motivated by the recent success of DL, an increasing number of studies have been performed to investigate the application of DL algorithms in the musculoskeletal system. Rajpurkar~\MakeLowercase{\textit{et al.}}~\cite{rajpurkar2017mura} proposed a DenseNet model~\cite{huang2017densely} to detect upper extremity abnormalities. In another study~\cite{urakawa2019detecting}, a VGG16-based model~\cite{simonyan2014very} was constructed to diagnose intertrochanteric hip fractures from proximal femoral radiographs, and the authors claimed their model exceeded the accuracy of orthopedic surgeons. However, the ROIs (i.e., proximal femurs) used to train the model were manually cropped from radiographs, which may hinder the potential application of the algorithm. To address this issue, Kazi~\MakeLowercase{\textit{et al.}}~\cite{kazi2017automatic} employed the spatial transformation network architecture~\cite{jaderberg2015spatial} to integrate tasks of the ROI localization and proximal femur fracture detection tasks into a single training process. Here, the diagnosis process can be automated, which broadens the scope of its potential application. As the ROIs were learned implicitly with the fracture detection training process, the performance and training efficiency of the entire model may not be optimal. Gale~\MakeLowercase{\textit{et al.}}~\cite{gale2017detecting} trained individual DCNN models using small subsets of data to achieve three pre-processing tasks, e.g., lateral-view detection and ROI localization. Thus, an individual model was constructed to focus on fracture diagnosis, and this model demonstrated state-of-the-art performance compared to previous methods.

In the AVNFH diagnosis context, Chee~\MakeLowercase{\textit{et al.}}~\cite{chee2019performance} trained a ResNet-based model~\cite{he2016deep} for AVNFH detection from 1,800 AP view pelvic radiographs with corresponding MRI as a data annotation reference. To integrate findings from both MRI and radiographs, the authors applied the Association Research Circulation Osseous (ARCO) guideline~\cite{gardniers1993arco} for AVNFH staging. A non-inferiority analysis was conducted and validated the performance of the DL algorithm is non-inferior to that of radiologists. However, as the pathological stage I of AVNFH is only visible on MRI, the authors did not clearly discuss how the MRI information was used to analyze radiographs. Moreover, using the AP view radiographs alone for AVNFH diagnosis is sometimes insufficient because subtle changes in the subchondral region may be missing owing to the overlapped anterior and posterior acetabular in the AP view~\cite{mont1995non}. In addition, all FHs were manually cropped from radiographs which would impact the utility of the model.

Differing from~\cite{chee2019performance}, in this study, we focus on radiographs from both AP and FL views to perform AVNFH diagnosis based on the Fiact system. Rather than training a single diagnosis model, we developed an end-to-end solution with state-of-the-art image classification and object detection algorithms to conduct the diagnostic process automatically, which substantially enhances performance and utility. Furthermore, we completed studies to assess the utility of the proposed AVN-net in two real-world application scenarios.

\subsection{Object detection with deep learning}
Nowadays, DL-driven object detection algorithms are widely used in various applications, e.g., self-driving car and security surveillance. Region-based CNNs (R-CNN) is a family of algorithms that represent a pioneering DL solution for object detection~\cite{girshick2014rich, girshick2015fast, ren2015faster}. However, computational efficiency is always the most significant concern for R-CNNs because they perform a classification computation for all extracted regions, and the number of candidate regions for a single input image can be up to 2,000. Then, the YOLO (i.e., You Only Look Once) object detection family was proposed to address the low-efficiency problem by splitting the image into a grid~\cite{redmon2016you, redmon2017yolo9000, redmon2018yolov3}. Besides, the single-shot multi-box detector (SSD) model was designed to balance the speed and precision trade-off by predicting categories and box offsets on different feature scales~\cite{liu2016ssd}. As a result, the SSD can efficiently detect bounding boxes of the target and perform classification in a single forward pass of the network. Moreover, RetinaNet~\cite{lin2017focal} adopts the same idea of feature pyramids as in SSD along with the novel focal loss to further improve the performance.

\begin{table*}[!t]
\caption{DATASET CHARACTERISTICS}
	\label{tab:data}
	\centering
	\resizebox{0.88\textwidth}{!}{
	    \begin{tabular}{lcl}
	    \hline
	    \multicolumn{1}{c}{\textbf{\begin{tabular}[c]{@{}c@{}}Datasets \&\\ Characteristics\end{tabular}}} & \textbf{\begin{tabular}[c]{@{}c@{}}AVNFH Staging (\# of FHs)\\ (Absence: Stage II: Stage III: Stage IV)\end{tabular}} & \multicolumn{1}{c}{\textbf{Uses}} \\ \hline
	    
	    \begin{tabular}[c]{@{}l@{}}\textbf{Total}\\~~~~841 subjects\\~~~~3,136 radiographs, 5,089 FHs\\~~~~AP: 2096, FL:1040\\ \end{tabular} & 1025 : 810 : 1271 : 1983 & For DL models training and testing \\
	    
	    \begin{tabular}[c]{@{}l@{}}\textbf{T-set}\\~~~~50 subject\\~~~~179 radiographs, 279 FHs\\ \end{tabular} & 52 : 47 : 59 : 121         & \begin{tabular}[c]{@{}l@{}}For performance testing of well-tuned DL models\end{tabular} \\
	    
	    \begin{tabular}[c]{@{}l@{}}\textbf{E1-set}\\~~~~100 subjects\\~~~~100 radiographs, 200 FHs\\ \end{tabular} & 38 : 61 : 61 : 40 & \begin{tabular}[c]{@{}l@{}}For performance comparison to orthopedists who performed diagnosis \\without referring the results of the proposed system (AVN-net).\end{tabular}   \\
	    
	    \begin{tabular}[c]{@{}l@{}}\textbf{E2-set}\\~~~~30 subjects\\~~~~30 radiographs, 60 FHs\\ \end{tabular} & 8 : 16 : 21 : 15 & \begin{tabular}[c]{@{}l@{}}For performance comparison to orthopedists who performed diagnosis \\with the help of AVN-net in the diagnosis support study (pilot A).\end{tabular} \\
	    
	    \begin{tabular}[c]{@{}l@{}}\textbf{E3-set}\\~~~~30 subjects\\~~~~30 radiographs, 60 FHs\\ \end{tabular} & 14 : 15 : 17 : 14 & \begin{tabular}[c]{@{}l@{}}For performance evaluation between two groups of students in\\ education assistance study (pilot B).\end{tabular}       \\ \hline
	    \end{tabular}
    }
\end{table*}

\section{Materials and Methods}

\subsection{Data collection and annotation}
All radiographs used in this study were extracted from the radiology repository at Peking University Third Hospital (PUTH), which is one of the largest tertiary referral hospitals in China. The PUTH research ethics board approved this study. Radiographs from both AP and FL views of all AVNFH patients seen at the orthopedics clinic between 2005 and 2019 were included. We excluded radiographs with primary hip osteoarthritis, secondary osteoarthritis caused by developmental dysplasia of the hip, avascular necrosis with pathological fracture, and internal fixation remaining in the FH. 

A panel of three orthopedic surgeons with at least 15 years of clinical and surgical experience was recruited for data annotation. When labeling FH bounding boxes, exam-views, and sides, each annotator was responsible for 1/3 of the data without overlapping, considering the complexity of these tasks was relatively low. When staging AVNFH, the annotators individually labeled all FHs, and a list of typical cases for each stage was recorded. Also, a short note for each FH was provided by the annotators, which describes the clinical findings that support their diagnosis. There are 3, 3, and 2 types of notes given by the annotators for stages II, III, and IV, respectively. For instance, notes for stage II include sclerotic change, cystic change, and crescent sign without FH flattening; subchondral flattening/collapse, FH deformation, and both for stage III; and FH and acetabular deformation, joint space stenosis for stage IV. In case of disagreements, consensus discussion and majority voting were performed to obtain a final decision. Because the pathological stage I cannot be observed on plain radiographs, we combined stages 0 and I as the AVNFH absence class. Thus, the classes for AVNFH staging include AVNFH absence, stage II, stage III, and stage IV. To facilitate performance evaluations, we further grouped stages II to IV as the AVNFH presence class and defined stage II as the pre-collapse class and stage III combined with stage IV as the post-collapse class.

In summary, a total number of 3,136 pelvic radiographs (2,096 AP view, 1,040 FL view, containing 5,089 FHs) from 841 subjects were included in this study. From which, the data of 50 subjects (including 179 radiographs, 279 FHs) were held out for models testing, denoted as T-set. Three additional sets of radiographs were collected for the evaluation tasks of compare-to-professionals, diagnosis support, and education assistance, respectively, denoted as E1-set, E2-set, and E3-set. The characteristics, AVNFH staging information, and uses of all datasets are summarized in Table~\ref{tab:data}.

\subsection{FH detection and post-processing}
We adopted the SSD algorithm~\cite{liu2016ssd} to construct the FH detection model. To emphasize localization, we simplified the classification task with binary classes (i.e., FH vs non-FH). The training objective is to minimize multi-box loss, which is a weighted summation of the localization loss (i.e., the smooth L1 loss~\cite{girshick2015fast} between the detected and the ground-truth boundary box parameters) and confidence loss (i.e., the binary cross-entropy loss of the FH classification). The localization loss ($\mathscr{L}_{loc}$), confidence loss ($\mathscr{L}_{conf}$), and multi-box loss ($\mathscr{L}_{MB}$) are formulated as follows:
\begin{equation}
\small
\mathscr{L}_{loc}(x,l,g) = \sum_{i,j}x_{ij}\text{smooth}_{L1}(l_i-g_j)
\end{equation}
\begin{equation}
\label{equ:2}
\small
\begin{split}
\mathscr{L}_{conf}(x,c) = &-\sum_{i,j}{ x_{ij}\log{c_i} +(1-x_{ij})\log{(1-c_i)}}
\end{split}
\end{equation}
\begin{equation}
\small
\mathscr{L}_{MB}(x,c,l,g) = \frac{1}{N}(\mathscr{L}_{conf}(x,c) + \alpha \mathscr{L}_{loc}(x,l,g))
\end{equation}
where $l_i$ is the $i$-th predicted boundary box coordinates, $g_j$ is the $j$-th ground-truth coordinates, $x_{ij}=\{0,1\}$ is the indicator for the $i$-th default box matched to the $j$-th ground-truth box, $c_i$ is the confidence score for predicting class $i$, $N$ is the number of matched boxes, and $\alpha$ is the weight of the localization loss.

With the detected boundary boxes, a post-processing strategy was designed to augment the FH data. Specifically, we introduced a scaling factor ($s$) and multiplied it to the side of the detected FH area. Therefore, each detected FH could be resampled multiple times by adjusting the value of $s$. In this study, we resampled each detected FH five times (ten times for typical cases) by linearly varying $s\in[0.85,1.05]$. As a result, a total of 23,225 FH images were obtained for training the downstream modules. At last, all resampled FHs were resized to $224\times224$ to fit the classification models.

\subsection{Classifications}
We defined exam-view and side identification, AVNFH diagnosis, and clinical note generation as individual classification tasks based on the detected FHs. In consideration of training efficiency, we uniformly adopted the 18-layer ResNet architecture to construct the backbone for each subtask. ResNet and its variants address the vanishing gradients problem by introducing the identity shortcut connection structure and have been widely adopted in various studies. The details of ResNet were thoroughly discussed elsewhere~\cite{he2016deep}. For exam-view and side classifications, we adopted cross-entropy loss as the cost function, of which form is similar to (\ref{equ:2}). For the AVNFH diagnosis module, we initialized the weights with those of the well-trained exam-view classification model to facilitate the training process. Moreover, we employed focal loss~\cite{lin2017focal} (denoted as $\mathscr{L}_{D}$) as the cost function to encourage the diagnosis model to put more focus on the relatively hard and misclassified examples. Formally, $\mathscr{L}_{D}$ is defined as follows:
\begin{equation}
\small
\label{equ:1}
\begin{split}
\mathscr{L}_{D}(x) = &-\frac{1}{N}\sum_{i=1}^{N}\alpha_i(1-x_i)^\gamma\log{(x_i)}
\end{split}
\end{equation}
where $\alpha_i$ is a weighting factor to balance different examples, $x_i$ is the $i$-th predicted probability, $\gamma$ is the focusing parameter. 

We framed clinical notes generation as a multi-label classification problem based on the textual descriptions provided by annotators (see examples in section III.A) and diagnosis results generated by the diagnosis model. Specifically, we categorized the textual notes into different classes and assigned each FH a multi-hot label. Then, we trained a multi-label classifier to generate all possible notes associated with each input FH. Here, the output probabilities of the diagnosis model were taken as the prior belief and were multiplied to the outputs of the notes classifier to get the rectified posterior probabilities.

\begin{figure*}[t]
	\centering
	\includegraphics[width=0.82\textwidth]{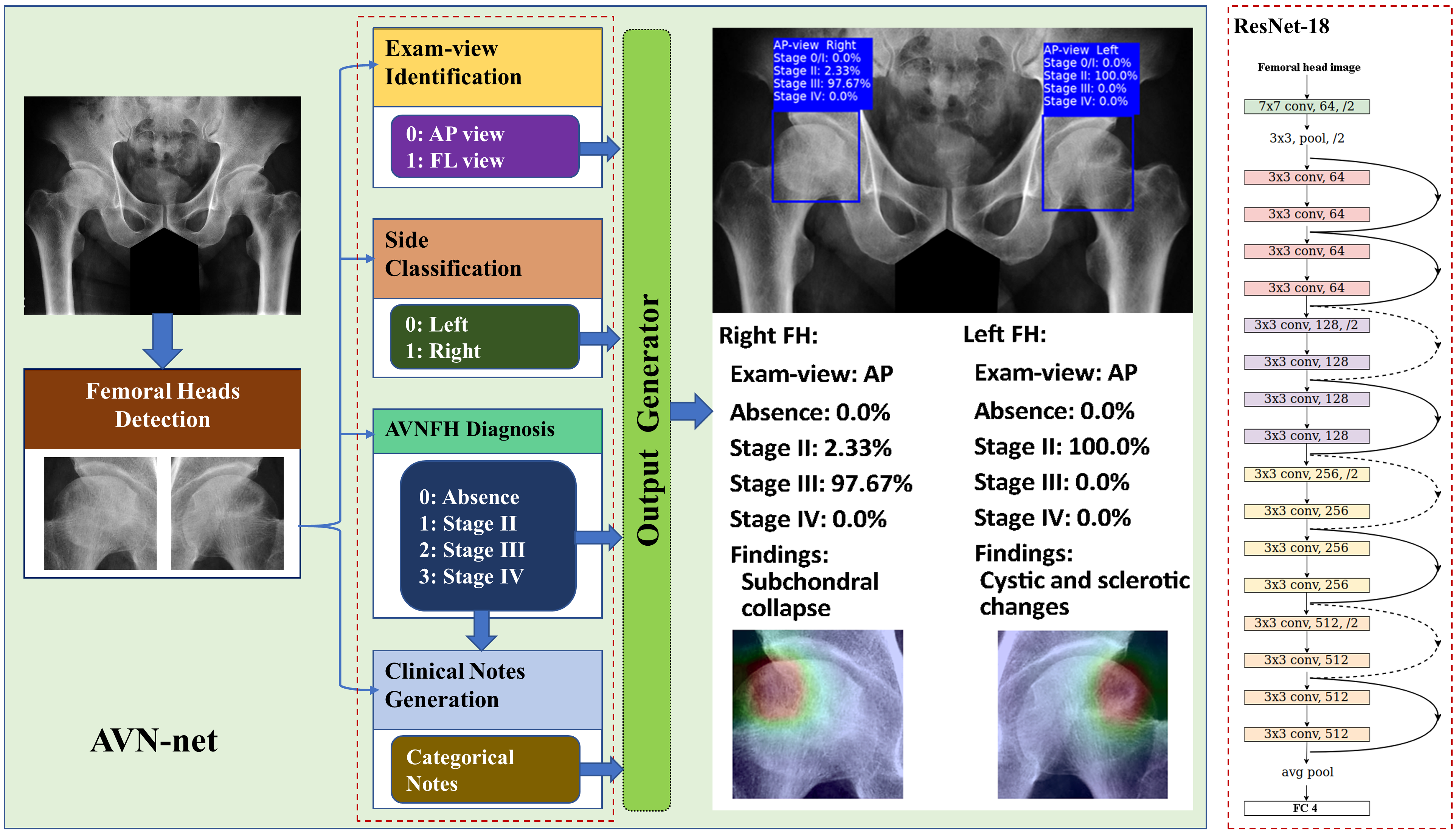} 
	\caption{System architecture: the FH detection module localizes FH boundaries from the input and passes them to downstream classification modules. Exam-view and side identification modules are binary classifiers. AVN diagnosis and clinical note generation modules are multi-class/label classifiers. All classifiers were structured with ResNet-18 model, of which architecture is shown in the red dashed box on the right. The output generator integrates results from all modules and visualizes the results. The class activation maps for visualization were generated based on~\cite{zhou2016learning}.}
	\label{fig:2}
\end{figure*}

\subsection{System architecture, models training and evaluation}
Figure~\ref{fig:2} shows the overall architecture of the proposed AVN-net. Pytorch~\cite{paszke2019pytorch} was used to implement all DL models. In the training phase, we performed 10-fold cross-validation (CV) to tune classification hyperparameters and validate models' performance on the training and validation sets. The data split ratio was set to 9:1 for the training and validation data in each fold. Standard data augmentation techniques were applied to the training data, including small-angle rotation, translation, horizontal flipping, etc. The initial learning rate for each task was set to 0.0001 and was halved when validation performance plateaus. We adopted Adam optimizer~\cite{kingma2014adam} with a weight decay of $0.0005$. After training, we tested the final classifiers on the T-set. Then, we repeated this CV-and-test procedure ten times, where, in each repetition of the 10-fold CV, we split folds differently by varying the random seed. The performance statistics of the classification models over repetitions are calculated. 

We assessed FH detection performance by measuring precision, recall, and F1, which are defined as follows:
\begin{equation}
\small
Precision = \frac{True~Positive}{True~Positive + False~Positive}
\end{equation}

\begin{equation}
\small
Recall = \frac{True~Positive}{True~Positive + False~Negative}
\end{equation}

\begin{equation}
\small
F1 = 2\cdot	\frac{Precision \times Recall}{Precision + Recall}
\end{equation}

The true positive (TP) and false positive (FP) cases were determined by measuring intersection over union (IoU). Explicitly, for each ground truth FH box, we defined the detection as a TP case if the corresponding IoU was greater than a predefined threshold. Otherwise, it was considered a FP. And, we counted the undetected FH as FN cases.

For classification models, we evaluated the AVN's performance on the T-set with AUC, sensitivity, specificity, F1 score, and corresponding 95\% confidence intervals (CI) over the ten times repeated CV. For the diagnosis model, the performance will be evaluated in three diagnostic sub-tasks, i.e., AVNFH detection (absences vs. presence), FH pre-/post-collapse classification (AVNFH absence vs. FH pre-collapse vs. post-collapse), and AVNFH staging. We conducted ablation analysis to investigate the influence of FH detection module and different resampling strategies on the performance. Furthermore, we compared the performance of our AVNFH staging model to that of four orthopedists with different experience levels on E1-set. The participants included two less-to-moderately experienced residents with average clinical experience of 3 years (denoted as group R), and two experienced attending surgeons with 7.5 years clinical experience on average (denoted as group S). None of these participants were involved in the data annotation process. The receiver operating characteristic (ROC) curves of AVN-net as well as the sensitivity and specificity of participating orthopedists in staging E1-set are plotted. The time required for each participant to complete evaluations was recorded.

\begin{table}[t]
	\caption{TEST PERFORMANCE OF FH DETECTION WITH DIFFERENT IOU THRESHOLDS}
	\label{tab:1}
	\centering
	\begin{tabular}{@{}cccc@{}}
		\toprule
		\textbf{IoU threshold} & \textbf{Precision} & \textbf{Recall} & \textbf{F1 score} \\ \midrule
		\textbf{0.5}           & \textbf{1.0}       & \textbf{0.9819} & \textbf{0.9908}   \\
		0.6                    & 0.9708             & 0.9815          & 0.9761            \\
		0.7                    & 0.9264             & 0.9805          & 0.9527            \\\bottomrule
	\end{tabular}
\end{table}

\begin{table}[t]
	\caption{TEST PERFORMANCE OF EXAM-VIEW, SIDE, AND CLINICAL NOTES CLASSIFICATION}
	\label{tab:2}
	\centering
	\resizebox{0.9\columnwidth}{!}{
		\begin{tabular}{@{}cccc@{}}
			\toprule
			\textbf{Task}& 
			\textbf{\begin{tabular}[c]{@{}c@{}}AUC\\ (95\% CI)\end{tabular}}  & \textbf{\begin{tabular}[c]{@{}c@{}}Sensitivity\\ (95\% CI)\end{tabular}} & \textbf{\begin{tabular}[c]{@{}c@{}}Specificity\\ (95\% CI)\end{tabular}} \\ 
			\midrule
			Side
			& \begin{tabular}[c]{@{}c@{}}0.9995 \\ (0.9987-1.0)\end{tabular}    & \begin{tabular}[c]{@{}c@{}}0.9965 \\ (0.9914-1.0)\end{tabular}           & \begin{tabular}[c]{@{}c@{}}0.9963 \\ (0.9939-0.9987)\end{tabular}           \\
			Exam-view
			& \begin{tabular}[c]{@{}c@{}}0.9997 \\ (0.9990-1.0)\end{tabular}    & \begin{tabular}[c]{@{}c@{}}0.9962 \\ (0.9935-0.9989)\end{tabular}           & \begin{tabular}[c]{@{}c@{}}0.9965 \\ (0.9947-0.9983)\end{tabular}           \\
			\begin{tabular}[c]{@{}c@{}}Clinical note \\ generation\end{tabular} & \begin{tabular}[c]{@{}c@{}}0.9637 \\ (0.9603-0.9671)\end{tabular} & \begin{tabular}[c]{@{}c@{}}0.8132 \\ (0.8038-0.8226)\end{tabular}        & \begin{tabular}[c]{@{}c@{}}0.9236 \\ (0.9174-0.9298)\end{tabular}        \\ \bottomrule
		\end{tabular}
	}
\end{table}

\begin{table}[t]
	\caption{PERFORMANCE COMPARISON TO STATE-OF-THE-ART RESULTS IN DIFFERENT TASKS}
	\label{tab:3}
	\centering
	\resizebox{\columnwidth}{!}{
		\begin{tabular}{@{}lcc@{}}
			\toprule
			\multicolumn{1}{c}{\textbf{Task and metrics}} & \textbf{Proposed AVN-net}         & \textbf{Chee \MakeLowercase{\textit{et al.}}} \\ \midrule
			\textbf{AVNFH detection}           & \multicolumn{1}{l}{}                                                      & \multicolumn{1}{l}{}                                                     \\
			~~\begin{tabular}[c]{@{}c@{}}AUC per FH~(95\% CI)\end{tabular}                                   & \begin{tabular}[c]{@{}c@{}}\textbf{0.974}~(0.971-0.978)\end{tabular} & \begin{tabular}[c]{@{}c@{}}0.930~(N/A)\end{tabular}            \\
			~~\begin{tabular}[c]{@{}c@{}}Sensitivity per FH~(95\% CI)\end{tabular}                                   & \begin{tabular}[c]{@{}c@{}}\textbf{0.946}~(0.937-0.955)\end{tabular} & \begin{tabular}[c]{@{}c@{}}0.848~(0.733-0.906)\end{tabular}            \\
			~~\begin{tabular}[c]{@{}c@{}}Specificity per FH~(95\% CI)\end{tabular}                                   & \begin{tabular}[c]{@{}c@{}}0.900~(0.877-0.923)\end{tabular} & \begin{tabular}[c]{@{}c@{}}\textbf{0.913}~(0.720-0.989)\end{tabular}            \\
			~~Sensitivity per subject~(95\% CI) & \multicolumn{1}{l}{} & \multicolumn{1}{l}{}   \\
			~~~~\begin{tabular}[c]{@{}c@{}}AP+FL view\end{tabular}                                   & \begin{tabular}[c]{@{}l@{}}\textbf{0.977}~(0.974-0.979)\end{tabular} & 
			\begin{tabular}[c]{@{}c@{}}N/A\end{tabular}            \\ 
			~~~~\begin{tabular}[c]{@{}c@{}}AP view\end{tabular}                                   & \begin{tabular}[c]{@{}c@{}}0.951~(0.941-0.961)\end{tabular} & 
			\begin{tabular}[c]{@{}c@{}}0.960~(0.888-0.986)\end{tabular} \\ 
			~~~~\begin{tabular}[c]{@{}c@{}}FL view\end{tabular}                                   & \begin{tabular}[c]{@{}c@{}}0.938~(0.927-0.948)\end{tabular} & 
			\begin{tabular}[c]{@{}c@{}}N/A\end{tabular} \\ 
			
			\textbf{FH pre-/post-collapse classification}    & \multicolumn{1}{l}{} & \multicolumn{1}{l}{} \\
			~~\begin{tabular}[c]{@{}c@{}}Sensitivity of pre-collapse~(95\% CI)\end{tabular}                  & \begin{tabular}[c]{@{}c@{}}\textbf{0.880}~(0.860-0.901)\end{tabular}      & \begin{tabular}[c]{@{}c@{}}0.759~(0.624-0.865)\end{tabular}            \\
			~~\begin{tabular}[c]{@{}c@{}}Sensitivity of post-collapse~(95\% CI)\end{tabular}                  & \begin{tabular}[c]{@{}c@{}}\textbf{0.932}~(0.919-0.946)\end{tabular}      & \begin{tabular}[c]{@{}c@{}}0.915~(0.825-0.965)\end{tabular}   \\ 
			\bottomrule
		\end{tabular}
	}
\end{table}

\begin{table}[t]
	\caption{TEST AVNFH STAGING PERFORMANCE}
	\label{tab:4}
	\centering
	\begin{threeparttable}
		\resizebox{\columnwidth}{!}{
			\begin{tabular}{@{}ccccc@{}}
				\toprule \multicolumn{1}{l}{}   & 
				\textbf{Absence}    & \textbf{Stage II} & \textbf{Stage III}    & \textbf{Stage IV}\\ \midrule
				\begin{tabular}[c]{@{}c@{}}AUC\\ (95\% CI)\end{tabular} & 
				\begin{tabular}[c]{@{}c@{}}0.9805\\ (0.978-0.982)\end{tabular} & 
				\begin{tabular}[c]{@{}c@{}}0.9284\\ (0.920-0.937)\end{tabular} & 
				\begin{tabular}[c]{@{}c@{}}0.9286\\ (0.921-0.936)\end{tabular} & 
				\begin{tabular}[c]{@{}c@{}}0.9923\\ (0.991-0.994)\end{tabular} \\
				Sensitivity$^{a}$   & 0.9651    & 0.8662    & 0.8070    & 0.9483    \\
				Specificity$^{b}$   & 0.8990    & 0.8586    & 0.8889    & 0.9697    \\ \bottomrule
			\end{tabular}
		}
		\begin{tablenotes}
			\item $^{a, b}$ Sensitivity and specificity values at the operating points.
		\end{tablenotes}
	\end{threeparttable}
\end{table}

\begin{table*}[t]
	\caption{AVNFH STAGING PERFORMANCE ABLATION ANALYSIS}
	\label{tab:10}
	\centering
	\begin{threeparttable}
		\resizebox{0.75\textwidth}{!}{
			\begin{tabular}{@{}ccccccc@{}}
				\toprule
				\multicolumn{1}{l}{\multirow{2}{*}{}}   & 
				\multirow{2}{*}{\textbf{\begin{tabular}[c]{@{}c@{}}Without FH\\ Detection\end{tabular}}} & 
				\multicolumn{5}{c}{\textbf{With FH Detection}} \\  
				\cmidrule(l){3-7} 
				\multicolumn{1}{l}{}    & & \textbf{N=2}$^{a}$    & \textbf{N=3}  & \textbf{N=5} 
				& \textbf{N=10}   & \textbf{N=20} \\ \midrule
				\begin{tabular}[c]{@{}c@{}}\textbf{F1 score}\\ (95\% CI)\end{tabular} & 
				\begin{tabular}[c]{@{}c@{}}0.7735\\ (0.7611-0.7859)\end{tabular}  & 
				\begin{tabular}[c]{@{}c@{}}0.7891\\ (0.7827-0.7955)\end{tabular} & 
				\begin{tabular}[c]{@{}c@{}}0.8032\\ (0.7949-0.8115)\end{tabular} & 
				\begin{tabular}[c]{@{}c@{}}\textbf{0.8369}\\ (0.8280-0.8458)\end{tabular} & 
				\begin{tabular}[c]{@{}c@{}}0.8368\\ (0.8270-0.8466)\end{tabular} & 
				\begin{tabular}[c]{@{}c@{}}0.8024\\ (0.7946-0.8102)\end{tabular} \\
				\textbf{p-value1}$^{b}$   & --    & $p$\textless{}0.05  & $p$\textless{}0.01  & $p$\textless{}0.001   
				& $p$\textless{}0.001 & $p$\textless{}0.01\\
				\textbf{p-value2}$^{c}$   & $p$\textless{}0.001 & $p$\textless{}0.001 & $p$\textless{}0.001
				& --    & $p$=0.99  & $p$\textless{}0.01    \\ 
				\bottomrule
			\end{tabular}
		}
		\begin{tablenotes}
			\item $^{a}$ N is the number of each detected FH was resampled with different scales.
			\item $^{b}$ T-test p-values on F1 scores between cases without and with FH detection.
			\item $^{c}$ T-test p-values on F1 scores between the proposed resample strategy (N=5) and others.
		\end{tablenotes}
	\end{threeparttable}
\end{table*}

\begin{table*}[t]
	\caption{ORTHOPEDISTS V.S AVN-NET}
	\label{tab:5}
	\centering
	\begin{threeparttable}
		\resizebox{0.83\textwidth}{!}{
			\begin{tabular}{@{}lccccc@{}}
				\toprule
				\multicolumn{1}{c}{\textbf{Tasks and Metrics}}  & \textbf{AVN-net}  & 
				\textbf{R1$^{a}$} & \textbf{R2}   & \textbf{S1}  & \textbf{S2}  \\ \midrule
				\begin{tabular}[c]{@{}l@{}}\textbf{AVNFH detection}\\
					~~~~F1 score (95\% CI)\\         ~~~~p-value$^{b}$\end{tabular}             & 
				\begin{tabular}[c]{@{}c@{}}\\0.9242 (0.8889-0.9646)\\ --\end{tabular} & 
				\begin{tabular}[c]{@{}c@{}}\\0.8586 (0.8131-0.9091)\\ $p$\textless 0.05\end{tabular} & 
				\begin{tabular}[c]{@{}c@{}}\\0.8131 (0.7626-0.8687)\\ $p$\textless{}0.01\end{tabular} & 
				\begin{tabular}[c]{@{}c@{}}\\0.9192 (0.8838-0.9596)\\ $p$=0.85\end{tabular}           & 
				\begin{tabular}[c]{@{}c@{}}\\\textbf{0.9293} (0.8939-0.9646)\\ $p$=0.84\end{tabular} \\
				\begin{tabular}[c]{@{}l@{}}\textbf{FH pre- vs. post-collapse}\\
					~~~~F1 score (95\% CI)\\~~~~p-value\end{tabular} & 
				\begin{tabular}[c]{@{}c@{}}\\0.8586 (0.8131-0.9091)\\ --\end{tabular} & 
				\begin{tabular}[c]{@{}c@{}}\\0.7576 (0.7020-0.8182)\\ $p$\textless{}0.01\end{tabular} & 
				\begin{tabular}[c]{@{}c@{}}\\0.7172 (0.6566-0.7828)\\ $p$\textless{}0.01\end{tabular} & 
				\begin{tabular}[c]{@{}c@{}}\\0.8636 (0.8182-0.9141)\\ $p$=0.88\end{tabular}           & 
				\begin{tabular}[c]{@{}c@{}}\\\textbf{0.8939} (0.8535-0.9394)\\ $p$=0.28\end{tabular} \\
				\begin{tabular}[c]{@{}l@{}}\textbf{AVNFH staging} \\~~~~F1 score (95\% CI)\\~~~~p-value\end{tabular}                       & 
				\begin{tabular}[c]{@{}c@{}}\\\textbf{0.8232} (0.7727-0.8788)\\ --\end{tabular} & 
				\begin{tabular}[c]{@{}c@{}}\\0.6263 (0.5606-0.6919)\\ $p$\textless{}0.01\end{tabular} & 
				\begin{tabular}[c]{@{}c@{}}\\0.6361 (0.5704-0.7117)\\ $p$\textless{}0.01\end{tabular} & 
				\begin{tabular}[c]{@{}c@{}}\\0.7768 (0.7273-0.8434)\\ $p$\textless{}0.01\end{tabular} & 
				\begin{tabular}[c]{@{}c@{}}\\0.8131 (0.7626-0.8687)\\ $p$=0.79\end{tabular} \\ \bottomrule
			\end{tabular}
		}
		\begin{tablenotes}
			\item This table shows the performance comparison between AVN-net and orthopedists with data of 10,000 bootstrap resampling in different tasks.
			\item $^{a}$ R1 and R2 represent two less-to-moderately experienced residents; S1 and S2 represent two experienced surgeons. 
			\item $^{b}$ T-test p-values on F1 scores between AVN-net and orthopedists in different tasks.
		\end{tablenotes}
	\end{threeparttable}
\end{table*}

\begin{figure*}[!t]
	\centering
	\begin{subfigure}[b]{0.255\textwidth}
		\includegraphics[width=\linewidth]{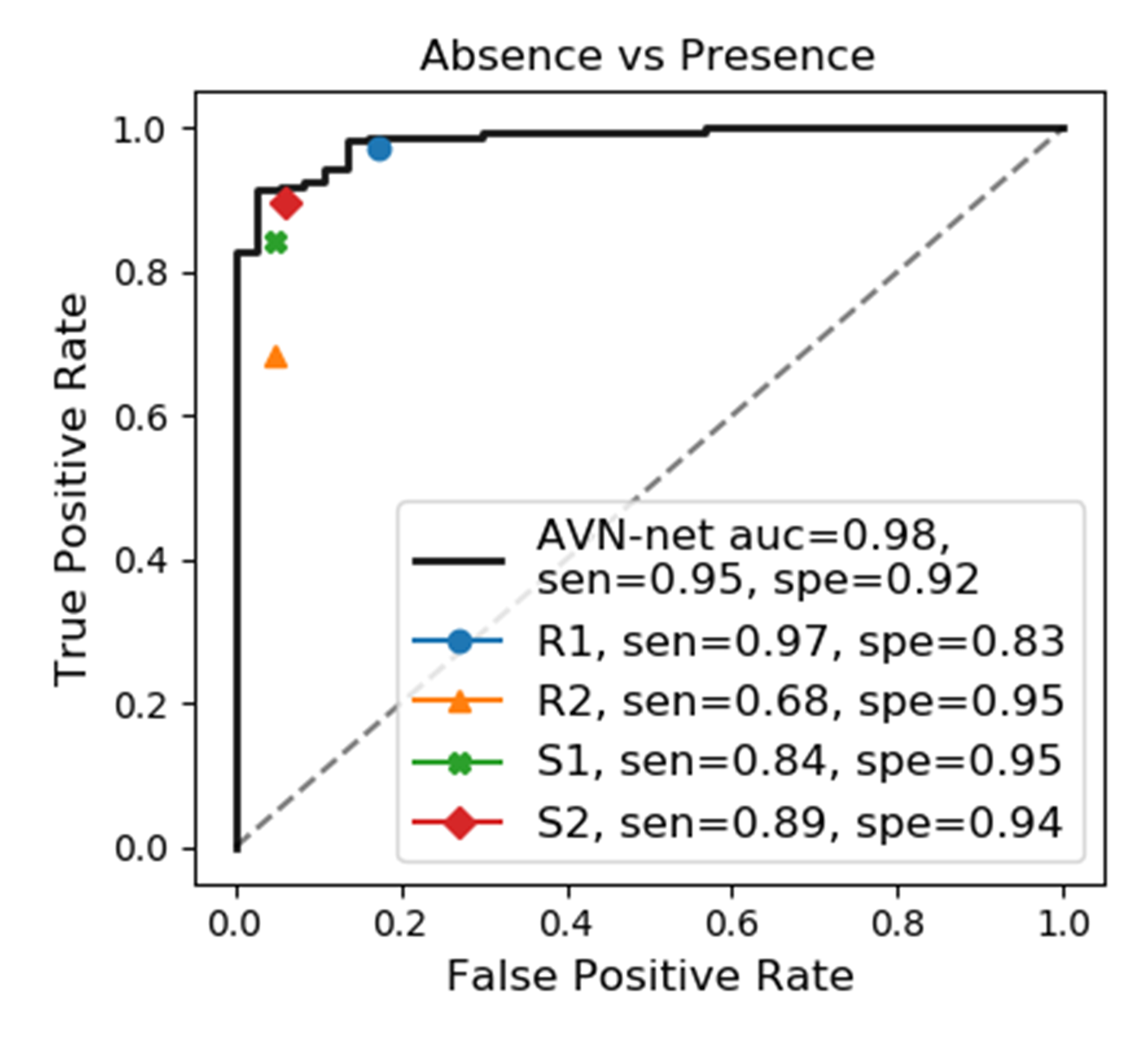}
		\caption{AVNFH detection on E1-set}
		\label{fig:3a}
	\end{subfigure}
	\begin{subfigure}[b]{0.46\textwidth}
		\includegraphics[width=\linewidth]{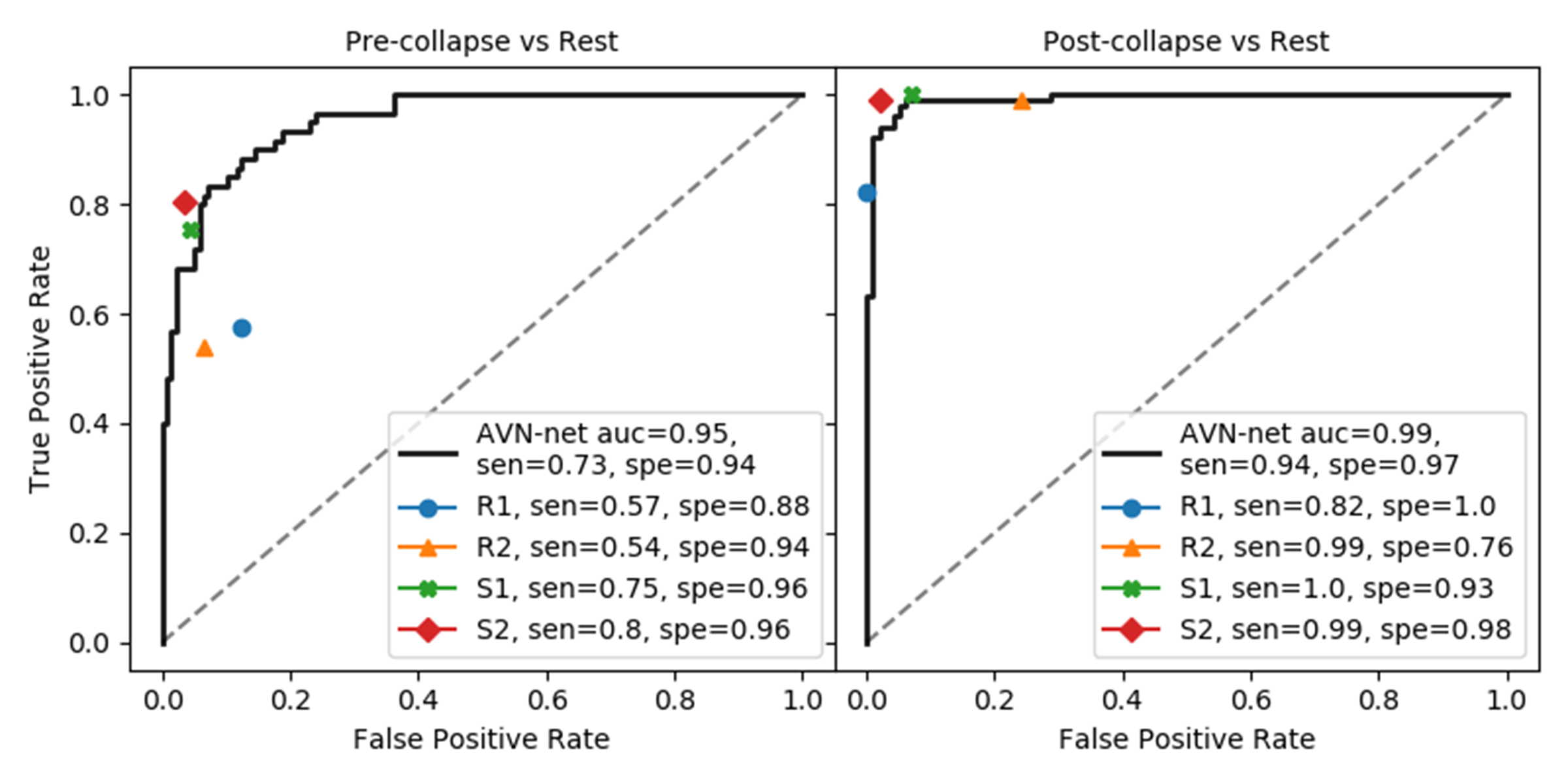}
		\caption{FH pre-/post-collapse classification on E1-set}
		\label{fig:3b}
	\end{subfigure}
	\begin{subfigure}[b]{0.91\textwidth}
		\includegraphics[width=\textwidth]{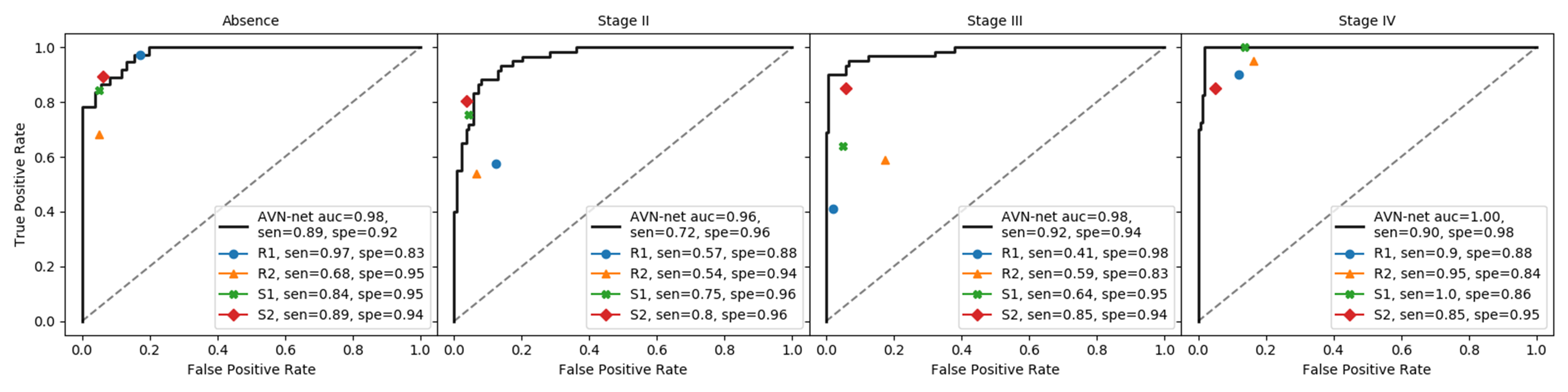}
		\caption{AVNFH staging on E1-set}
		\label{fig:3c}
	\end{subfigure}
	\begin{subfigure}[b]{0.91\textwidth}
		\includegraphics[width=\textwidth]{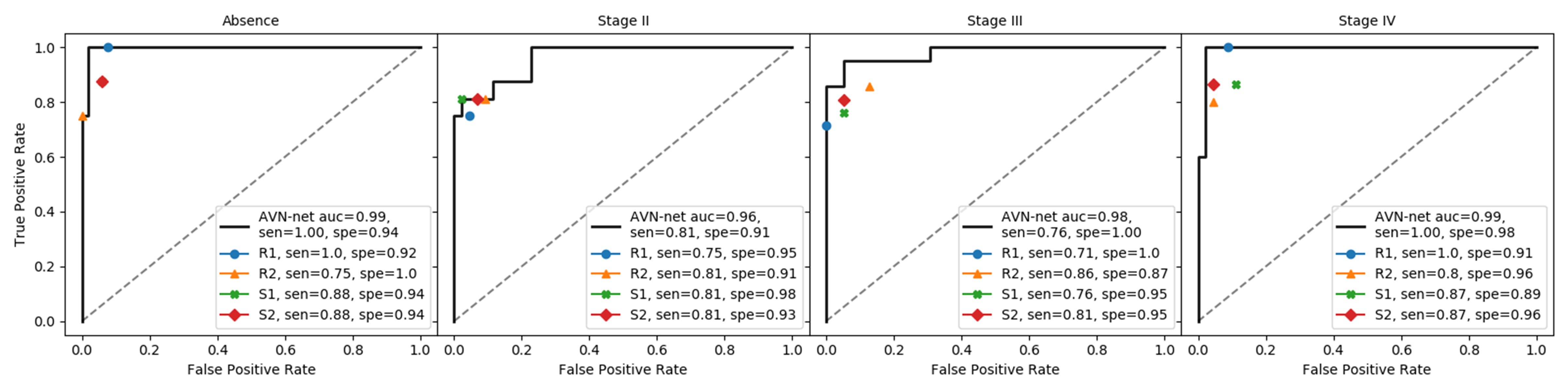}
		\caption{AVNFH staging on E2-set}
		\label{fig:3d}
	\end{subfigure}
	\caption{Diagnostic performance in different evaluation tasks. (a) AVNFH detection performance on E1-set (orthopedists vs. AVN-net). (b) Pre-/post-collapse classifications performance on E1-set (orthopedists vs. AVN-net). (c) AVNFH staging performance on E2-set (orthopedists vs. AVN-net). (d) AVNFH staging performance on E2-set (orthopedists collaborated with AVN-net). Note: Sen and spe in legends represent sensitivity and specificity.}
	\label{fig:3}
\end{figure*}

\section{Results}
For the FH detection task, as the performance plateaus when decreasing the IoU threshold to a particular value, to balance the performance and quality of the detected FHs trade-off, we empirically set the IoU threshold as the first value that gives the precision of 1, i.e. 0.5. In this case, the FH detection model achieved precision, recall, and F1 scores of 1.0, 0.9819, and 0.9908 on T-set. Table~\ref{tab:1} lists the performance of the FH detection model with different IoU threshold values. 

For the side classification task, the mean AUC on T-set was 0.9995 (95\% CI, 0.9987-1.0) with sensitivity and specificity values of 0.9965 (95\% CI, 0.9914-1.0) and 0.9963 (95\% CI, 0.9939-0.9987). The exam-view classifier obtained a mean testing AUC of 0.9997 (95\% CI, 0.9990-1.0) with sensitivity of 0.9962 (95\% CI, 0.9935-0.9989) and specificity of 0.9965 (95\% CI, 0.9947-0.9983). For the clinical note generation task, the proposed model obtained a mean AUC of 0.9637 (95\% CI, 0.9603-0.9671) with the mean sensitivity and specificity values of 0.8132 (95\% CI, 0.9038-0.9226) and 0.9236 (95\% CI, 0.9174-0.9298), respectively. The performance of corresponding tasks is summarized in Table~\ref{tab:2}.

When testing the AVNFH detection model on FH level, i.e., classifying AVNFH absence vs. presence on each FH image, AVN-net obtained a mean testing AUC of 0.9735 on T-set (95\% CI, 0.9707-0.9782). When performing AVNFH detection on the subject level, AVN-net examines all radiographs associated with the same subject and takes the most severe stage as the final diagnosis for the subject. In this case, AVN-net achieved the sensitivity of 0.9767 (95\% CI, 0.9739-0.9794) on radiographs from both AP and FL exam views. Note that the sensitivity of using two-view radiographs is significantly greater ($p$\textless0.01) than that of using single view: either AP view (0.9512, 95\% CI 0.9413-0.9611) or FL view (0.9375, 95\% CI 0.9272-0.9479). In the FH pre-/post-collapse classification task, AVN-net attained a mean testing AUC of 0.9371 (95\% CI, 0.9330-0.9412) for the pre-collapse classification, and a mean testing AUC of 0.9624 (95\% CI, 0.9594-0.9654) for the post-collapse classification. Table~\ref{tab:3} shows that the proposed AVN-net outperforms the state-of-art results (Chee~\MakeLowercase{\textit{et al.}}) in both evaluation sub-tasks. For the AVNFH staging sub-task, the proposed AVN-net obtained a mean testing AUC of 0.9805 (95\% CI: 0.9783-0.9824), 0.9284 (95\% CI: 0.9200-0.9368), 0.9286 (95\% CI: 0.9058-0.9514), and 0.9923 (95\% CI: 0.9911-0.9935) for AVNFH absence, Stage II, Stage III, and Stage IV classes, respectively. The detailed performance for AVNFH staging model is given in Table~\ref{tab:4}. 

To understand the contribution of the FH detection module to the diagnostic performance, we examined the mean F1 score of the AVNFH staging model trained without FH detection (i.e., cropping from radiographs with predefined raw ROIs) and models trained with different resampling strategies (i.e., using other resampling target numbers). Table~\ref{tab:10} shows that the AVNFH staging models with FH detection significantly outperform the model without FH detection ($p$\textless0.05), regardless of the number of resampled FHs. Resampling each detected FH five times (in various scales) obtains the best performance.

When evaluating the performance of AVN-net against that of orthopedists on the E1-set, AVN-net obtained sensitivity and specificity of 0.9543 and 0.9231 in the AVNFH detection task. In contrast, the mean sensitivity and specificity for groups R and S were 0.8253 and 0.8898, and 0.8601 and 0.9461. For FH pre-collapse classification, the sensitivity and specificity were 0.7333 and 0.9420 for AVN-net, 0.5548 and 0.9120 for group R, and 0.7755 and 0.9627 for group S. For post-collapse classification, the sensitivity and specificity for AVN-net were 0.9406 and 0.9691, 0.9057 and 0.8821 for group R, and 0.9953 and 0.9576 for group S. For AVNFH staging, the sensitivity and specificity values obtained by AVN-net for the AVNFH absence, stage II, stage III, and stage IV classes were 0.8919 and 0.9193, 0.7167 and 0.9565, 0.9180 and 0.9416, 0.9000 and 0.9810, respectively. Here, group R obtained the mean sensitivity and specificity values of 0.8289 and 0.8889, 0.5573 and 0.9065, 0.5008 and 0.9029, and 0.9250 and 0.8594, respectively. Group S obtained the mean sensitivity and specificity values of 0.8684 and 0.9444, 0.7786 and 0.9604, 0.7459 and 0.9460, and 0.9257 and 0.9067, respectively. Figures~\ref{fig:3a}-\ref{fig:3c} illustrate the ROC curves of AVN-net and the performance of orthopedists in the three tasks. In addition, the average time required to diagnose each radiograph for groups R and S was 44.75s and 23.04s, respectively.

We applied bootstrapping with replacement to construct the CI of F1 scores for AVN-net and orthopedists on E1-set. We calculated F1 scores for 10,000 bootstrapping samples in each sub-task, then took the $2.5^{th}$ and $97.5^{th}$ percentiles of the differences between the observed and bootstrapping F1 scores to estimate its 95\% CI for corresponding sub-tasks. Table~\ref{tab:5} shows that AVN-net obtained a significantly higher F1 score ($p$\textless0.05) than the less-to-moderately experienced orthopedists (group R) for all three sub-tasks and significantly better F1 ($p$\textless0.01) than one experienced surgeon in the AVNFH staging task, while maintaining the same performance level as the experienced group for the remaining tasks.

\begin{table*}[t]
	\caption{PERFORMANCE OF ORTHOPEDISTS IN COLLABORATIVE DIAGNOSIS STUDY}
	\label{tab:6}
	\centering
	\begin{threeparttable}
		\resizebox{0.85\textwidth}{!}{
			\begin{tabular}{@{}cccccc@{}}
				\toprule
				\multicolumn{2}{c}{\textbf{Metrics}} & \textbf{R1}                                                                          & \textbf{R2}    & \textbf{S1}   & \textbf{S2}   \\ \midrule
				
				\multirow{3}{*}{\begin{tabular}[c]{@{}c@{}}\textbf{F1 score} \\(95\% CI)\\~\end{tabular}}  & 
				w/o$^{a}$ AVN-net   & 0.6263 (0.5606-0.6919)  & 0.6361 (0.5404-0.6717)  & 0.7768 (0.7273-0.8434)  & 0.8131 (0.7626-0.8687)  \\& 
				w/ AVN-net    & 0.8167 (0.7167-0.8833)  & 0.8333 (0.7333-0.9000)   & 0.8175 (0.7246-0.9254)   & 0.8333 (0.7413-0.9088)  \\&
				p-value$^{b}$ & $p$\textless{0.01} & $p$\textless{0.01} & $p$=0.49 & $p$=0.69\\
				\\
				
				\multirow{2}{*}{\begin{tabular}[c]{@{}c@{}}\textbf{Time$^{c}$}\\ (s/radiograph)\end{tabular}}  & 
				w/o AVN-net & 46.9  & 42.0  & 20.0  & 26.1  \\& 
				w/ AVN-net   & 14.3  & 8.6   & 5.2   & 7.1   \\ \bottomrule
			\end{tabular}
		}
		\begin{tablenotes}
			\item This table compares the F1 scores and time consumption of each orthopedist with and without AVN-net results on the data of 10,000 bootstrapping samples.  
			\item $^{a}$ w/o and w/ represent without and with the result of AVN-net for reference, respectively. 
			\item $^{b}$ T-test p-values on F1 scores for each orthopedist diagnosing with vs. without the use of AVN-net.
			\item $^{c}$ The average time consumption for staging one radiograph.
		\end{tablenotes}
	\end{threeparttable}
\end{table*}

\section{Pilot studies}
To assess the utility of AVN-net, we conducted two pilot studies to investigate how well AVN-net helps orthopedists and medical students with diagnosis and self-study.

\subsection{AVN-net in diagnosis support: collaborative diagnosis}
In this study, we invited the same group of orthopedists in previous experiments to diagnose 30 new radiographs (E2-set). All annotators confirm that E2-set have similar difficulty for diagnosis as E1-set. Here, participants were given all AVN-net results for reference when reading the radiographs, which included the diagnosis, class activation maps, and generated notes (e.g., the output of Fig.~\ref{fig:2}). Then, the diagnostic performance and time consumption of each orthopedist were calculated and compared to their previous records.

For group R, the mean sensitivity and specificity when diagnosing with AVN-net (without AVN-net) were 0.8750 and 0.9615 (0.8289 and 0.8889), 0.7813 and 0.9318 (0.5573 and 0.9065), 0.7857 and 0.9359 (0.5008 and 0.9029), 0.9000 and 0.9333 (0.9250 and 0.8594) for the AVNFH absence, stage II, stage III, and stage IV classes, respectively. For group S, the sensitivity and specificity with AVN-net (without AVN-net) were 0.8750 and 0.9423 (0.8684 and 0.9444), 0.8125 and 0.9545 (0.7786 and 0.9604), 0.7857 and 0.9487 (0.7459 and 0.9460), and 0.8777 and 0.9222 (0.9257 and 0.9067) for the AVNFH absence, stage II, stage III, and stage IV classes, respectively. Figure~\ref{fig:3d} shows the ROC curves of AVN-net in staging E2-set and orthopedists' performance collaborating with AVN-net. The mean sensitivity of detecting stages II and III, which were difficult tasks for orthopedists in previous experiments, improved by 48\% and 5\% for groups R and S, when collaboratively diagnosing using AVN-net.

We applied the same bootstrapping to construct the 95\% CI of F1 scores for each orthopedist in this study. Table~\ref{tab:6} shows that, by referring to the ``second opinion" from AVN-net, the F1 scores of all orthopedists were improved, and the improvements for both residents in group R were statistically significant ($p$\textless0.01). Regarding time consumption, group R spent an average of 11.45s reading each radiograph with AVN-net (44.45s without AVN-net), and group S spent 6.12s on average (23.05s without AVN-net), which were approximately four times faster than without AVN-net for both groups.

\begin{table}[t]
	\caption{TESTING PERFORMANCE OF TWO GROUPS IN EDUCATION ASSISTANCE STUDY}
	\label{tab:7}
	\centering
	\resizebox{0.93\columnwidth}{!}{
	\begin{threeparttable}
		\begin{tabular}{@{}ccc@{}}
			\toprule & 
			\textbf{\begin{tabular}[c]{@{}c@{}}Experimental\\ group\end{tabular}} & \textbf{\begin{tabular}[c]{@{}c@{}}Control\\ group\end{tabular}} \\ 
			\midrule
			\begin{tabular}[c]{@{}c@{}}\textbf{F1 score} (95\% CI)\\ p-value$^a$ \end{tabular} & \begin{tabular}[c]{@{}c@{}}\textbf{0.6750} (0.5833-0.7667)\\ $p$=0.35 \end{tabular} & 
			\begin{tabular}[c]{@{}c@{}}0.6167 (0.5333-0.7000) \\--\end{tabular} \\
			\\
			\begin{tabular}[l]{@{}l@{}}\textbf{Time} (s/radiograph)\end{tabular} & 70 & 75                                                               \\ 
			\bottomrule
		\end{tabular}
		
		\begin{tablenotes}
			\item $^a$ T-test p-value on F1 between experimental and control groups.  
		\end{tablenotes}
	\end{threeparttable}
	}
\end{table}

\subsection{AVN-net in education assistance}
In the field of medical education, learning from clinical experience plays an essential role in helping students assimilate knowledge from books and develop practical skills. However, due to limited educational resources, it is unrealistic for students to receive constant guidance from domain experts. Consequently, the quality of self-study is critical in medical education. To this end, we hypothesize that, in conjunction with knowledge learned from experienced specialists, AVN-net can provide instructive and practical help to students in self-study than textbooks alone. Therefore, we recruited four senior medical students with similar academic performance and clinical experience to learn AVNFH diagnosis in self-study to assess the effectiveness of the proposed AVN-net.

The students were randomly divided into control and experimental groups. A tutoring package, consisting of 40 radiographs from E1-set, the ground truth diagnosis, and a list of literature related to AVNFH and the Ficat classification system, was presented to both groups. The control group was asked to learn AVNFH staging with the given resources, and the experimental group was provided access to the AVN-net web application, with which they could observe the AVN-net diagnosis by uploading the provided examples or additional radiographs. Also, both groups were allowed to obtain additional information, e.g., from the Internet or textbooks, except for getting help from human experts. After a two-hour learning period, we tested both groups on E3-set. All radiographs in E3-set were typical cases for corresponding stages. The performance and time required by each student was also recorded and compared.

For the control group, the mean testing sensitivity and specificity for detecting AVNFH absence, stage II, stage III, stage IV were 0.4286 and 0.9457, 0.7333 and 0.8000, 0.6471 and 0.8488, and 0.6429 and 0.8913, respectively. For the experimental group, these values were 0.6071 and 0.9674, 0.7333 and 0.8222, 0.6765 and 0.6372, and 0.6786 and 0.9348. We constructed the 95\% CI of F1 scores for both groups using the same bootstrapping strategy. Table~\ref{tab:7} shows that the experimental group obtained a mean testing F1 score of 0.6750 (95\% CI, 0.5833-0.7667) which was greater (but not significantly greater) than that of the control group (0.6167, 95\% CI: 0.5333-0.7000). The mean time consumption for the experimental group was slightly less than that of the control group (70 vs. 75 s/radiograph).

\begin{table}[t]
	\caption{INTRA/INTER-READER AGREEMENT IN AVNFH STAGING}
	\label{tab:8}
	\centering
	\begin{threeparttable}
		\resizebox{0.86\columnwidth}{!}{
			\begin{tabular}{@{}ccccc|cccc@{}}
				\toprule
				\multirow{2}{*}{} & \multicolumn{4}{c|}{\textbf{Without AVN-net}} & \multicolumn{4}{c}{\textbf{With AVN-net}} \\ \cmidrule(l){2-9} 
				& R1     & R2     & S1     & S2    & R1    & R2    & S1    & S2    \\ \midrule
				R1                & 1      & 0.52   & 0.53   & 0.57  & 1     & 0.64  & 0.71  & 0.78  \\
				R2                & --     & 1      & 0.58   & 0.65  & --    & 1     & 0.68  & 0.68  \\
				S1                & --     & --     & 1      & 0.69  & --    & --    & 1     & 0.75  \\
				GT$^{a}$          & 0.56   & 0.55   & 0.71   & 0.79  & 0.78  & 0.72  & 0.75  & 0.77  \\ \bottomrule
			\end{tabular}
		}
		
		\begin{tablenotes}
			\item This table lists Cohen's Kappa statistics between each pair of orthopedists in ordinary and collaborative diagnosis modes.
			\item $^a$ GT represents ground truth labels.
		\end{tablenotes}
	\end{threeparttable}
\end{table}

\begin{figure}[t]
	\centering
	\resizebox{0.83\columnwidth}{!}{
		\includegraphics{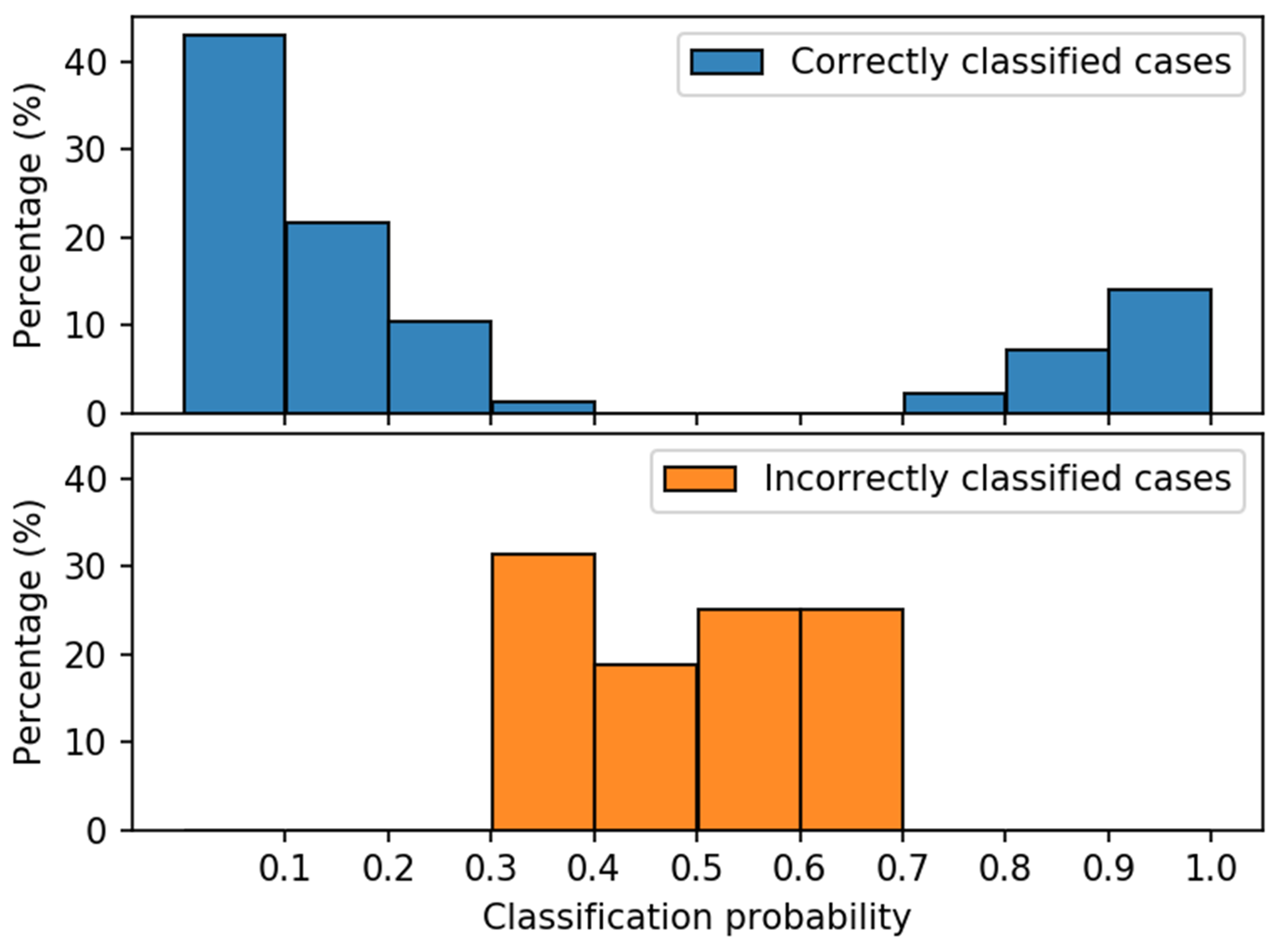} 
	}
	\caption{Uncertainty analysis: distributions of the classification probability for correctly and incorrectly classified samples}
	\label{fig:4}
\end{figure}

\section{Discussion}
As the first diagnostic imaging modality for AVNFH, it would be extremely beneficial for patient care if detect early-stage AVNFH from the radiograph in a timely manner. However, due to the modest changes between the stages of AVNFH, accurate diagnosis from a plain radiograph is a significant challenge, especially for relatively less experienced clinicians, which results in poor diagnostic consistency.

From previous experiments, we have confirmed that the proposed AVN-net is able to achieve state-of-the-art performance in various diagnostic tasks and significantly outperforms less-to-moderately experienced orthopedists. By referring to AVN-net, all orthopedists would obtain the same level of diagnostic performance as the domain expert, and the time required to examine each radiograph was reduced by a factor of four. 

To further investigate the influence of AVN-net on diagnostic consistency, we compared Cohen's Kappa coefficients~\cite{cohen1960coefficient} between orthopedists with and without AVN-net classes for reference in AVNFH staging. Table~\ref{tab:8} shows that, when collaborating with AVN-net, the intra-group diagnosis agreements for groups R and S (i.e., Kappa between readers in the same group) increased by 21.1\% and 8.7\%, and the inter-group agreement (i.e., the mean Kappa between the groups R and S) was improved by over 22.4\% (from 0.58 to 0.71). Consistency between group R and ground truth increased by 35.1\% when collaborating with AVN-net (from 0.555 to 0.75), and the value for group S was maintained at the same level (0.75 vs. 0.76). Therefore, we can get that AVN-net is especially valuable for less-to-moderately experienced orthopedists in diagnostic decision making. Also, when looking at cases that AVN-net misclassified, we found that, for each of such example, at least one orthopedist was able to make a correct diagnosis using AVN-net's suggestion, which, to some extent, proves that AVN-net’s second opinion cannot place human assessment.

In addition to assessing performance metrics, it is equally important to examine the classification uncertainty since it affects how users interpret the result. To this end, we respectively analyzed the distribution of classification probabilities on samples of T-set that were correctly and incorrectly classified by the AVNFH staging model. In Figure~\ref{fig:4}, the x-axis stands for the model's output probabilities, and the y-axis represents the percentage of samples corresponding to each decile. We can see those classifications assigned a high certainty by our diagnosis model (i.e., the output probability was either close to 0 or 1) were more likely correct; whereas, the incorrectly classified cases tend to have a low confidence (i.e., the output probability was close to 0.5). It confirms that our model behaves in a way that is consistent with our expectations. Note that AVNFH staging is a multi-classes task. To count the number of correct/incorrect cases, we treated each stage as an individual binary classification with the ``one vs. the rest" labels, which explains why the proportion of true-negative cases was much greater than true-positives.

\begin{figure}[t]
	\centering
	\resizebox{0.88\columnwidth}{!}{
		\includegraphics{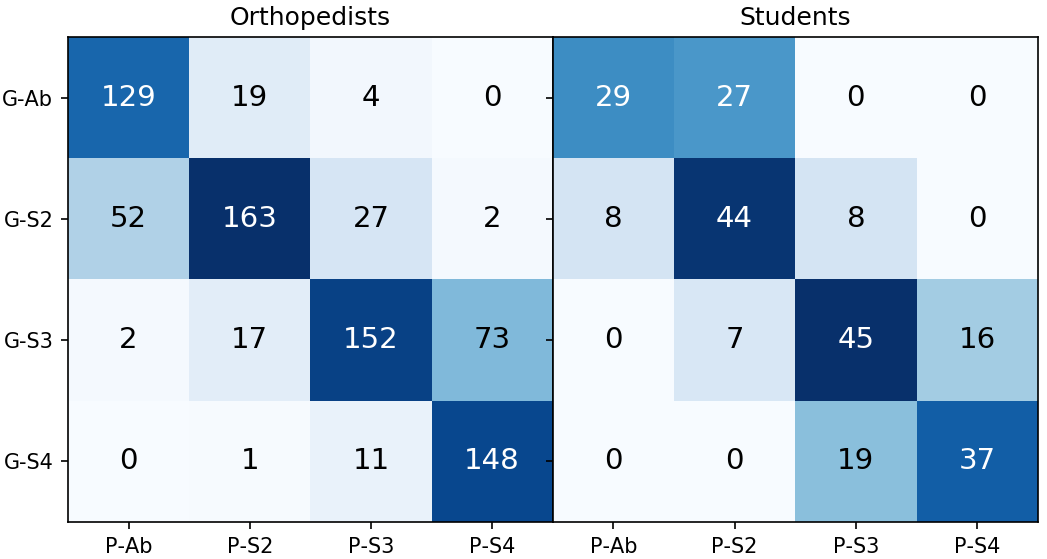} 
	}
	\caption{Confusion matrices of all orthopedists on E1-set and students on E3-set (performance without AVN-net)}
	\label{fig:5}
\end{figure}

We also obtained a promising result from the medical education study. The group of students who were self-learning using AVN-net achieved better test performance and spent slightly less time on average than the group of students that did not use AVN-net. In addition, we found that both groups obtained similar performance as the less experienced orthopedists. We think this could have resulted from the different number of radiographs and time spent on staging each radiograph, which also implies two possible causes of poor inter-reader agreement. When comparing the results of all students to that of orthopedists on E3 and E1 sets (i.e., to compare results without involving AVN-net), we found some distinct performance patterns. Fig.~\ref{fig:5} illustrates the confusion matrices of the two groups, where G/P in the label of each row/column represents the ground truth/predictions, Ab, S2, S3, and S4 represent the four classes, and the number in each cell is the number of predictions for the corresponding category. It can be seen that students were more sensitive to identifying stages II and III, which were difficult for the orthopedists, especially for the less-to-moderately experienced group. In contrast, the orthopedists achieved much higher accuracy in detecting extreme categories (i.e., AVNFH absence and stage IV) than the students. We suspect this can be attributed to the different clinical experience between the groups, which also suggests that different strategies should be designed for different users and application scenarios.

Due to a data limitation, we defined the clinical note generation task as a classification problem based on categorical descriptions. In future, we plan to take the diagnosis report as an additional data dimension because we believe the textual information could be beneficial to improve model explainability and performance. Also, we solely focused on AVNFH diagnosis from plain radiographs, and AVNFH stage I is not visible in X-ray images; thus, the capability of early-stage detection with AVN-net may be constrained. In the next step, we plan to extend this study's scope by involving CT or MRI data and subdivisions in each stage. In addition, as Gal~\MakeLowercase{\textit{et al.}} pointed out that Softmax activation has flaws in representing classification confidence~\cite{gal2016dropout}, it would be worthwhile to explore other means to measure model uncertainty, which is crucial for healthcare applications.

\section{Conclusion}
In this paper, we presented a fully automated end-to-end AVNFH diagnosis system based on DL algorithms and plain radiographs of the pelvis (AVN-net). The proposed AVN-net demonstrated state-of-the-art performance in various diagnostic tasks and significantly outperformed the less-to-moderately experienced orthopedists while maintaining same performance level of experienced orthopedists. We further performed two pilot studies to discuss the utility of the proposed AVN-net, where AVN-net was used for diagnosis support and education assistance. The results of both studies confirm that AVN-net can provide efficient and effective support in making diagnostic decisions and is especially helpful for less or moderately experienced orthopedic practitioners. 

\ifCLASSOPTIONcaptionsoff
  \newpage
\fi

\bibliographystyle{IEEEtran}
\bibliography{ref}

\end{document}